\documentstyle{article}

\title{Labels for Non-Individuals}
\author{Adonai S. Sant'Anna\thanks{Permanent address:
Department of Mathematics, Federal University of Paran\'a, P. O.
Box 019081, Curitiba, PR, 81531-990. E-mail: adonai@ufpr.br.}}

\date{Department of Philosophy\\University of South Carolina\\Columbia, SC, 29208}
\begin{document}

\newtheorem{definicao}{Definition}
\newtheorem{teorema}{Theorem}
\newtheorem{lema}{Lemma}
\newtheorem{corolario}{Corolary}
\newtheorem{proposicao}{Proposition}
\newtheorem{axioma}{Axiom}
\newtheorem{observacao}{Observation}

\maketitle


\begin{abstract}

Quasi-set theory is a first order theory without identity, which
allows us to cope with non-individuals in a sense. A weaker
equivalence relation called ``indistinguishability'' is an
extension of identity in the sense that if $x$ is identical to $y$
then $x$ and $y$ are indistinguishable, although the reciprocal is
not always valid. The interesting point is that quasi-set theory
provides us a useful mathematical background for dealing with
collections of indistinguishable elementary quantum particles. In
the present paper, however, we show that even in quasi-set theory
it is possible to label objects that are considered as
non-individuals. We intend to prove that individuality has nothing
to do with any labelling process at all, as suggested by some
authors. We discuss the physical interpretation of our results.

\end{abstract}

\section{Introduction}

Concerning non-individuality in quantum mechanics, the problems
raised by this have provided many papers in the literature. See,
for example, the references in French (2004).

Elementary particles that share the same set of state-independent
(intrinsic) properties are some times said to be {\em
indistinguishable\/}. Although classical particles can share all
their intrinsic properties, we might say that they `have' some
kind of {\em quid\/} which makes them individuals. Hence, we are
able to follow the trajectories of classical particles, at least
in principle. That allows us to identify them. In quantum physics
this is not possible, i.e., it is not possible, {\em a priori\/},
to keep track of individual particles in order to distinguish
among them when they share the same set of intrinsic properties.
In other words, it is not possible to label quantum particles. And
this non-individuality plays a very important role in quantum
mechanics (Sakurai, 1994).

On the possibility that collections of such indistinguishable
entities should not be considered as sets in the usual sense,
Manin (1976) proposed the search for axioms which should allow to
deal with indiscernible objects. As he said,

\begin{quote}
I would like to point out that it [standard set theory] is rather
an extrapolation of common-place physics, where we can distinguish
things, count them, put them in some order, etc. New quantum
physics has shown us models of entities with quite different
behavior. Even {\em sets\/} of photons in a looking-glass box, or
of electrons in a nickel piece are much less Cantorian than the
{\em sets\/} of grains of sand.
\end{quote}

We are using the philosophical jargon in saying that
`indistinguishable' objects are objects that share their
properties, while `identical' objects means `the very same
object'. Nevertheless, in considering the behavior of the
ensembles of such particles, there is a fundamental difference
between classical and quantum statistics. In classical statistical
mechanics, particles are treated like individuals. In quantum
statistics, on the other hand, the Indistinguishability Postulate
asserts that if a permutation is applied to any state for an
assembly of particles, then there is no way of distinguishing the
resulting permuted state-function from the original one by means
of any observation at any time. The Indistinguishability Postulate
(IP) seems to be one of the most basic principles of quantum
theory and implies that permutations of quantum particles are not
regarded as an observable.

Usually, IP has been interpreted in two basic ways: the first
assumes that IP implies that quantum particles cannot be regarded
as `individuals', since an `individual' should be something having
properties similar to those of usual (macroscopic) bodies. This
interpretation is closely related to what is assumed in the
context of quantum field theory, since, roughly speaking, quantum
field theories do not deal with `individuals' . The second way
regards particles as individuals in a sense, and the non-classical
counting of quantum statistics are then viewed as resulting from
the restrictions imposed to the set of the possible states
accessible to the particles. In short, only symmetrical and
anti-symmetrical states are available, and the initially attached
individuality of particles is then `veiled' by such a criterion.
Both alternatives, albeit used in current literature, present
problems from the `foundational' point of view. There is some
obscurity lurking in the concept of individuality in quantum
physics. The idea of considering `non-individuals' is weird, and
in general other metaphysical packages are used instead. For
instance, that one which assumes that quantum objects are
individuals of a sort, despite quite distinct from the usual
objects described by classical mechanics (Sant'Anna and Krause,
1997).

Let us recall that some authors like Hermann Weyl expressed the
calculation with `aggregates' so that some of the basic
assumptions of quantum theory can be reached in an adequate way.
Weyl's efforts were done in the sense of finding an alternative
manner to express the procedure physicists implicitly use in
treating indistinguishable particles, namely, the assumption that
there is a {\it set\/} $S$ of (hence, distinguishable) objects
(say, $n$ objects) endowed with an equivalence relation $\sim$.
Then the `desired result', according to Weyl, is to obtain the
{\it ordered decomposition\/} $n = n_{1} + \cdots + n_{k}$, where
$n_{i}$ are the cardinalities of the equivalence classes $C_{i}$,
$i = 1, \ldots, k$ of the quotient set $S/\sim$. But, as it is
easy to note, this procedure `veils' the very nature of the
elements of the set $S$, that is, veils the fact that they are
individual objects since they are members of a {\it set\/}. We
would like to emphasize that there is no scape. Classical logic
and mathematics are committed with a conception of identity which
does not make any distinction between identity and
indistinguishability: indistinguishable things are the very same
thing and conversely.

One manner to cope with the problem of non-individuality in
quantum physics is by means of quasi-set theory (Krause, 1992;
Krause, Sant'Anna, and Volkov, 1999; Sant'Anna and Santos, 2000),
which is an extension of Zermelo-Fraenkel set theory, that allows
to talk about certain indistinguishable objects that are not
identical. Such indistinguishable objects are termed as
non-individuals. In quasi-set theory identity does not apply to
all objects. In other words, there are some kinds of terms in
quasi-set theory where the sequence of symbols $x = y$ is not a
well-formed-formula, i.e., it is meaningless. A weaker equivalence
relation called ``indistinguishability'' is an extension of
identity in the sense that it allows the existence of {\em two\/}
objects that are indistinguishable. In standard mathematics, there
is no sense in saying that two objects are identical. If $x = y$,
then we are talking about one single object with two labels,
namely, $x$ and $y$.

We want to continue our investigations on the use of quasi-set
theory in the foundations of quantum mechanics, based on some
questions that we think about and that may be interesting. Some of
these questions have to do with the notion of {\em levels of
individuality\/}, which is introduced in further details in the
next sections. Actually, our main mathematical framework is some
sort of quasi-set-theoretical predicate for quantum systems, which
is a natural extension of Patrick Suppes (2002) ideas about
axiomatization. We prove, e.g., that even in quasi-set theory it
is possible to prove that objects without individuality (in the
sense of the theory) may be labelled if certain conditions are
satisfied. We want to investigate the meaning of this labelling
process from the point of view of formal logic and we want to
study the possibility of a new kind of quasi-set theory where this
kind of labelling process cannot be performed.

\section{Quasi-sets}

This section is strongly based on other works about quasi-set
theory (Krause, 1992; Krause, Sant'Anna, and Volkov, 1999;
Sant'Anna and Santos, 2000). I use standard logical notation for
first-order theories (Mendelson, 1997).

Quasi-set theory ${\cal Q}$ is based on Zermelo-Fraenkel-like
axioms and allows the presence of two sorts of atoms ({\it
Urelemente\/}), termed $m$-atoms (micro-atoms) and $M$-atoms
(macro-atoms). Concerning the $m$-atoms, a weaker `relation of
indistinguishability' (denoted by the symbol $\equiv$), is used
instead of identity, and it is postulated that $\equiv$ has the
properties of an equivalence relation. The predicate of equality
cannot be applied to the $m$-atoms, since no expression of the
form $x = y$ is a formula if $x$ or $y$ denote $m$-atoms. Hence,
there is a precise sense in saying that $m$-atoms can be
indistinguishable without being identical. This justifies what we
said above about the `lack of identity' to some objects.

The universe of ${\cal Q}$ is composed by $m$-atoms, $M$-atoms and
{\it quasi-sets\/} (qsets, for short). The axiomatization is
adapted from that of ZFU (Zermelo-Fraenkel with {\it
Urelemente\/}), and when we restrict the theory to the case which
does not consider $m$-atoms, quasi-set theory is essentially
equivalent to ZFU, and the corresponding quasi-sets can then be
termed `sets' (similarly, if also the $M$-atoms are ruled out, the
theory collapses into ZFC). The $M$-atoms play the same role of
the {\it Urelemente\/} in the sense of ZFU.

In all that follows, $\exists_Q$ and $\forall_Q$ are the
quantifiers relativized to quasi-sets. That is, $Q(x)$ reads as
`$x$ is a quasi-set'.

In order to preserve the concept of identity for the
`well-behaved' objects, an {\it Extensional Equality\/} is defined
for those entities which are not $m$-atoms on the following
grounds: for all $x$ and $y$, if they are not $m$-atoms, then $$x
=_{E} y := \forall z ( z \in x \Leftrightarrow z \in y ) \vee
(M(x) \wedge M(y) \wedge x \equiv y).$$

It is possible to prove that $=_{E}$ has all the properties of
classical identity in a first order theory and so these properties
hold regarding $M$-atoms and `sets'. In this text, all references
to `$=$' (in quasi-set theory) stand for `$=_E$', and similarly
`$\leq$' and `$\geq$' stand, respectively, for `$\leq_E$' and
`$\geq_E$'. Among the specific axioms of ${\cal Q}$, few of them
deserve explanation. The other axioms are adapted from ZFU.

For instance, to form certain elementary quasi-sets, such as those
containing `two' objects, we cannot use something like the usual
`pair axiom', since its standard formulation assumes identity; we
use the weak relation of indistinguishability instead:

The `Weak-Pair' Axiom - For all $x$ and $y$, there exists a
quasi-set whose elements are the indistinguishable objects from
either $x$ or $y$. In symbols,

$$\forall x \forall y \exists_{Q} z \forall t (t \in z
\Leftrightarrow t \equiv x \vee t \equiv y).$$

Such a quasi-set is denoted by $[x, y]$ and, when $x \equiv y$, we
have $[x]$, by definition. We remark that this quasi-set {\it
cannot\/} be regarded as the `singleton' of $x$, since its
elements are {\it all\/} the objects indistinguishable from $x$,
so its `cardinality' (see below) may be greater than $1$. A
concept of {\it strong singleton\/}, which plays a crucial role in
the applications of quasi-set theory, may be defined.

In ${\cal Q}$ we also assume a Separation Schema, which
intuitively says that from a quasi-set $x$ and a formula
$\alpha(t)$, we obtain a sub-quasi-set of $x$ denoted by $$[t\in x
: \alpha(t)].$$

We use the standard notation with `$\{$' and `$\}$' instead of
`$[$' and `$]$' only in the case where the quasi-set is a {\it
set\/}.

It is intuitive that the concept of {\it function\/} cannot also
be defined in the standard way, so we introduce a weaker concept
of {\it quasi-function\/}, which maps collections of
indistinguishable objects into collections of indistinguishable
objects; when there are no $m$-atoms involved, the concept is
reduced to that of function as usually understood. Relations (or
{\em quasi-relations\/}), however, can be defined in the usual
way, although no order relation can be defined on a quasi-set of
indistinguishable $m$-atoms, since partial and total orders
require antisymmetry, which cannot be stated without identity.
Asymmetry also cannot be supposed, for if $x \equiv y$, then for
every relation $R$ such that $\langle x, y \rangle \in R$, it
follows that $\langle x, y \rangle =_{E} [[x]] =_{E} \langle y, x
\rangle \in R$, by force of the axioms of ${\cal Q}$.

We remark that $[[x]]$ is the same ($=_{E}$) as $\langle x, x
\rangle$ by the Kuratowski's definition.

It is possible to define a translation from the language of ZFU
into the language of ${\cal Q}$ in such a way that we can obtain a
`copy' of ZFU in ${\cal Q}$. In this copy, all the usual
mathematical concepts (like those of cardinal, ordinal, etc.) can
be defined; the `sets' (in reality, the `${\cal Q}$-sets' which
are `copies' of the ZFU-sets) turn out to be those quasi-sets
whose transitive closure (this concept is like the usual one) does
not contain $m$-atoms.

Although some authors like Weyl (1949) sustain that (in what
regard cardinals and ordinals) ``the concept of ordinal is the
primary one'', quantum mechanics seems to present strong arguments
for questioning this thesis, and the idea of presenting
collections which have a cardinal but not an ordinal is one of the
most basic and important assumptions of quasi-set theory.

The concept of {\it quasi-cardinal\/} is taken as primitive in
${\cal Q}$, subject to certain axioms that permit us to operate
with quasi-cardinals in a similar way to that of cardinals in
standard set theories. Among the axioms for quasi-cardinality, we
mention those below, but first we recall that in ${\cal Q}$,
$qc(x)$ stands for the `quasi-cardinal' of the quasi-set $x$,
while $Z(x)$ says that $x$ is a {\it set\/} (in ${\cal Q}$).
Furthermore, $Cd(x)$ and $card(x)$ mean `$x$ is a cardinal' and
`the cardinal of $x$', respectively, defined as usual in the
`copy' of ZFU.

Quasi-cardinality - Every qset has an unique quasi-cardinal which
is a cardinal (as defined in the `ZFU-part' of the theory) and, if
the quasi-set is in particular a set, then this quasi-cardinal is
its cardinal {\em stricto sensu}:

$$\forall_{Q} x \exists_{Q} ! y (Cd(y) \wedge y =_{E} qc(x) \wedge
(Z(x) \Rightarrow y =_{E} card(x))).$$

Then, every quasi-cardinal is a cardinal and the above expression
`there is a unique' makes sense. Furthermore, from the fact that
$\emptyset$ is a set, it follows that its quasi-cardinal is 0
(zero).

${\cal Q}$ still encompasses an axiom which says that if the
quasi-cardinal of a quasi-set $x$ is $\alpha$, then for every
quasi-cardinal $\beta \leq \alpha$, there is a sub-quasi-set of
$x$ whose quasi-cardinal is $\beta$, where the concept of {\it
sub-quasi-set\/} is like the usual one. In symbols,

The quasi-cardinals of sub-quasi-sets - $$\forall_{Q} x (qc(x)
=_{E} \alpha \Rightarrow \forall \beta (\beta \leq_{E} \alpha
\Rightarrow \exists_{Q} y (y \subseteq x \wedge qc(y) =_{E}
\beta)).$$

Another axiom states that

The quasi-cardinal of the power quasi-set -
$$\forall_{Q} x (qc({\cal P}(x)) =_{E} 2^{qc(x)}).$$

\noindent
where $2^{qc(x)}$ has its usual meaning.

As remarked above, in ${\cal Q}$ there may exist qsets whose
elements are $m$-atoms only, called `pure' qsets. Furthermore, it
may be the case that the $m$-atoms of a pure qset $x$ are
indistinguishable from one another, in the sense of sharing the
indistinguishability relation $\equiv$. In this case, the
axiomatization provides the grounds for saying that nothing in the
theory can distinguish among the elements of $x$. But, in this
case, one could ask what it is that sustains the idea that there
is more than one entity in $x$. The answer is obtained through the
above mentioned axioms (among others, of course). Since the
quasi-cardinal of the power qset of $x$ has quasi-cardinal
$2^{qc(x)}$, then if $qc(x) = \alpha$, for every quasi-cardinal
$\beta \leq \alpha$ there exists a sub-quasi-set $y \subseteq x$
such that $qc(y) = \beta$, according to the axiom about the
quasi-cardinality of the sub-quasi-sets. Thus, if $qc(x) = \alpha
\not= 0$, the axiomatization does not forbid the existence of
$\alpha$ sub-quasi-sets of $x$ which can be regarded as
`singletons'.

Of course the theory cannot prove that these `unitary'
sub-quasi-sets (supposing now that $qc(x) \geq 2$) are distinct,
since we have no way of `identifying' their elements, but qset
theory is compatible with this idea. In other words, it is
consistent with ${\cal Q}$ to maintain that $x$ has $\alpha$
elements, which may be regarded as absolutely indistinguishable
objects. Since the elements of $x$ may share the relation
$\equiv$, they may be further understood as belonging to a same
`equivalence class' (for instance, being indistinguishable
electrons) but in such a way that we cannot assert either that
they are identical or that they are distinct from one another
(i.e., they act as `identical electrons' in the physicist's
jargon).

We define $x$ and $y$ as {\it similar\/} qsets (in symbols,
$Sim(x,y)$) if the elements of one of them are indistinguishable
from the elements of the other one, that is, $Sim(x,y)$ if and
only if $\forall z \forall t (z \in x \wedge t \in y \Rightarrow z
\equiv t)$. Furthermore, $x$ and $y$ are {\it Q-Similar\/}
($QSim(x,y)$) if and only if they are similar and have the same
quasi-cardinality. Then, since the quotient qset $x/_{\equiv}$ may
be regarded as a collection of equivalence classes of
indistinguishable objects, then the `weak' axiom of extensionality
is:

Weak Extensionality -
\begin{eqnarray}
\forall_{Q} x \forall_{Q} y (\forall z (z \in x/_{\equiv}
\Rightarrow \exists t (t \in y/_{\equiv} \wedge \, QSim(z,t))
\wedge \forall t(t \in
y/_{\equiv} \Rightarrow\nonumber\\
\exists z (z \in  x/_{\equiv} \wedge \, QSim(t,z)))) \Rightarrow x
\equiv y)\nonumber
\end{eqnarray}

In other words, this axiom says that those qsets that have `the
same quantity of elements of the same sort (in the sense that they
belong to the same equivalence class of indistinguishable objects)
are indistinguishable.

\section{Some Applications}

Quasi-set theory has found its way in the sense of some
applications in quantum physics. Here we list some of them:

\begin{enumerate}

\item It has been used (Krause, Sant'Anna, and Volkov, 1999) for
an authentic proof of the quantum distributions. By ``authentic
proof'' we mean a proof where elementary quantum particles are
really considered as non-individuals right at the start. If the
physicist says that some particles are indistinguishable (in a
sense) and he/she still uses standard mathematics in order to cope
with these particles, then something seems not to be sound. For
standard mathematics is based on the concept of individuality, in
the sense that it is grounded on the very notion of identity.

\item It has been proved (Sant'Anna and Santos, 2000) that even
non-individuals may present a classical distribution like
Maxwell-Boltzmann's. That is another way to say that a
Maxwell-Boltzmann distribution in an ensemble of particles does
not entail any ontological character concerning such particles, as
it was previously advocated by Nick Huggett (1999).

\item Krause, Sant'Anna, and Volkov (1999) also introduced the
quasi-set-theoretical version of the wave-function of the atom of
Helium, which is a well known example where indistinguishability
plays an important role. Other discussions may be found in the
cited reference.

\end{enumerate}

\section{Individualizing Indiscernible Objects}\label{algoritmo}

This section presents the main contribution of the present paper.
We introduce an algorithm which allows us to label indiscernible
objects in the context of quasi-set theory. The algorithm is given
below, followed by its interpretation and discussion.

\begin{description}
\item[1.] INPUT $[ x ]$
\item[2.] DO $m =_E 0$
\item[3.] DO $w =_E \emptyset$
\item[4.] DO $m := m + 1$
\item[5.] DO $[ x ] := [ x ] - x'$
\item[6.] DO $w := w\cup [\langle x', m\rangle]$
\item[7.] OUTPUT $w$
\item[8.] IF $[ x ] =_E \emptyset$ THEN GO TO {\bf 10}
\item[9.] GO TO {\bf 4}
\item[10.] END
\end{description}

In the first step, we introduce a finite weak singleton $[ x ]$,
i.e., a pure quasi-set with a finite quasi-cardinality (a finite
number of elements) where all its elements are indistinguishable
objects. Next, we introduce a variable $m$ with an initial value
equal to zero and another variable which is an empty quasi-set
$w$. In the fourth step we transform $m$ into $m + 1$ (we use as
attribution symbol the sign ``$:=$''). In the fifth step we
subtract one of the elements of qset $[ x ]$ by means of a
difference between the weak singleton $[ x ]$ and the strong
singleton $x'$ (a weak singleton whose elements are
indistinguishable from $x$ but such that $x'$ has actually just
{\em one\/} element). Next we create an ordered pair defined by
$x'$ and by the label $m$. In step seven, this ordered pair is an
output. Actually the ordered pairs are stored in $w$. In a sense,
this works as a data warehousing process. Such process repeats
from step {\bf 4} until step {\bf 7} up to the moment when the
qset $[ x ]$ is empty.

So, in a sense, it is possible to `individualize' (by means of
integer labels $m$) objects that have no individuality in
principle. We are seriously tempted to refer to micro-atoms as
non-individuals, since the standard identity does not apply to
them. Nevertheless, there is nothing within the scope of quasi-set
theory that forbids us to attribute labels to micro-atoms, even in
a finite collection of indistinguishable micro-atoms.

From the physical point of view this is like to say that two
electrons, separated by a large distance between them, may be
identified by their coordinates in spacetime. Another
interpretation is the next one. Consider that we have a $Na$ atom,
with its 11 electrons. Consider also that we excite this atom by
adding a new electron to it. When the atom emits an electron,
there is no way to know if the emitted electron is the `same' one
used to excite the $Na$ atom. In this sense, the electrons of the
$Na$ atom are indistinguishable, they are non-individuals. But
when an electron is emitted, this particle may be labelled as that
one which was emitted. For some very restrictive purposes, it is
possible to label even non-individuals. But the most important
point is that this labelling process is possible even in a formal
mathematical framework which was designed specially for
applications in quantum theories, namely, quasi-set theory.

So, we advocate that we need to be very careful with respect to
the notions of identity and indistinguishability.

\section{A Conclusion}

What is individuality? If we answer this, it seems reasonable to
consider, at least in principle, that the notion of
non-individuality is also settled. That seems to happen because we
intend to consider that a given object is a non-individual if it
is not an individual at all. In order to talk about the
individuality of an object, do we need to talk about its
properties, like suggested by some authors (French, 2004)? Or
should the individuality of an object be expressed in terms of its
'haecceity' or 'primitive thisness' like suggested by Adams
(1979). Actually, I am not concerned with these points in this
paper. Otherwise, a much longer discussion should be made. The
point that I want to emphasize has to do with the possible
relationship between individuality and the labelling process.
Dalla Chiara and Toraldo di Francia (1993), for example, refer to
quantum physics as 'the land of anonymity', in the sense that,
particles cannot be uniquely labelled. Many other authors have a
similar opinion. But I want to advocate a different position on
this.

Quasi-set theory is a set theory without identity that allows the
existence of collections of objects that are indistinguishable in
the sense of the properties of the equivalence relation $\equiv$.
In some cases, this indistinguishability collapses to the usual
identity. That means that, in some cases, when we say that
$x\equiv y$, then we are talking about just {\em one\/} object,
named $x$ and $y$. It is usual to consider that such objects are
individuals due to their uniqueness. Nevertheless, quasi-set
theory allows also the existence of some objects (some
micro-atoms) such that $x\equiv y$ does not ensure that we are
really talking about one single (or unique) object. Any object $x$
that is indistinguishable from $y$ in this sense cannot be an
individual, since it lacks the notion of uniqueness. Nevertheless,
physicists usually consider that elementary particles are not
individuals in the sense that they cannot be labelled and we
cannot keep track of them in spacetime.

But with our algorithm we have proved that even objects that are
considered as non-individuals right at the start can be labelled,
at least in a weak sense. By weak sense me mean that two ordered
pairs of $w$ can always be distinguished by means of the second
element of the pair; but, on the other hand, the first coordinates
of these two pairs are still indistinguishable in the
quasi-set-theoretical sense. So, we think that we have proved (at
least in the context of quasi-set theory) that individuality have
nothing to do with any labelling process. The fact that we can
label some object does not guarantee that this object is an
individual in a precise sense.

Obviously, there is a limitation in our algorithm. Since this is a
step-by-step algorithm, it works only for finite or denumerable
weak singletons $[ x ]$. By denumerable weak singleton we mean a
qset $[ x ]$ whose quasi-cardinality is the cardinality of the set
of natural numbers. But for any physical interpretation of $[ x
]$, only finite weak singletons make sense. And my concern here is
with the notion of individuality (or non-individuality) in
physics, with special emphasis on quantum physics.

The ability of labelling non-individuals makes perfect sense from
the physical point of view. See, e.g., the Einstein-Podolsky-Rosen
Gedanken experiment (Sakurai, 1994), where two distant electrons
seem to be strongly connected by some sort of non-local
`interaction'. In this experiment the two electrons are formally
considered as indistinguishable, due to their entangled quantum
state. Nevertheless, someone could say that these electrons can be
labelled by their coordinates in the three dimensional space. But
the fact that these electrons can be labelled by their space
coordinates does not affect the fact that they are truly
indistinguishable, i.e., that they are non-individuals. A similar
situation happens with electrons in an atom. If an atom has 11
electrons, all of them are considered as non-individuals, although
they can be labelled by their respective quantum states, obeying
Pauli's Principle of Exclusion.

\section{Open Problems}

Here we present two lists of open problems that we consider worth
of investigation. Our main goal, in the present Section, is to
propose some ideas for future papers related to quasi-set theory
and the problems of non-individuality in quantum mechanics.

The first list has to do with our labelling algorithm.

\begin{enumerate}

\item Is it possible to create some kind of quasi-quasi-set theory
where the algorithm introduced in section \ref{algoritmo} does not
work? The existence of strong singletons is a crucial aspect in
the algorithm. Another point is that the membership relation $\in$
in quasi-set theory is like the usual one. What should we change
in quasi-set theory in order to avoid the labelling process of our
algorithm? And if this new theory is possible, then it makes sense
to talk about ``levels of individuality''. One first level would
be in correspondence with the standard view of identity in first
order and higher order theories; a second level would be in
correspondence to the notion of indistinguishability in quasi-set
theory as presented here; and a third level of individuality would
be a very weak form where there would exist some kind of
indistinguishability such that our algorithm does not apply.

\item If it is possible to create some sort of quasi-quasi-set
theory where our algorithm does not work, another question is: can
we use this quasi-quasi-set theory in order to ground the
foundations of quantum mechanics? Can we derive, e.g., the quantum
statistics in this new framework? That would be a way for a better
understanding of the meaning of non-individuality among quantum
particles.

\end{enumerate}

Our second list concerns related problems about quasi-set theory.

\begin{enumerate}

\item Maxwell-Boltzmann statistics can be derived even in a
collection of indiscernible particles (Sant'Anna and Santos,
2000). This means that Maxwell-Boltzmann statistics is not
necessarily committed to individuality. What are the physical (or
metaphysical) implications of this kind of result?

\item How to derive the spin-statistics theorem (Ryder, 1996)
within the scope of quasi-set theory?

\item Some authors have considered that there is a close
relationship between indistinguishability and non-locality in
quantum theory. One of them is Mandel (1991), who achieved some
interesting results in the context of the two-slit experiment. On
the other hand Einstein-Podolsky-Rosen (EPR) {\em Gedanken\/}
experiment is also one of those outstanding results that have
allowed many interesting discussions about the concept of
non-locality. We speculate if it is possible to extend the notion
of metric space in order to replace the notion of identity by the
concept of indistinguishability in the axioms of distance, since
nonlocal phenomena occur only between indistinguishable particles
(or indistinguishable paths). In metric spaces it is considered
that if $p=q$ then the distance between $p$ and $q$ is zero. What
if we consider that the distance is zero if and only if $p\equiv
q$? In this case, nonlocal phenomena would be more reasonable,
since the distance (in this new metric space) between
indistinguishable particles would be always null. The
philosophical issue here is to consider that space and spacetime
are concepts that are closely related to physical systems in the
sense that space and spacetime are derived concepts and not
primitive ones (da Costa and Sant'Anna, 2001; da Costa and
Sant'Anna, 2002).

\end{enumerate}

\section{Acknowledgements}

I would like to thank the important suggestions and criticisms
made by D\'ecio Krause and Newton C. A. da Costa.

I would like to thank Ot\'avio Bueno and Davis Baird for their
hospitality during my stay at the Department of Philosophy of the
University of South Carolina.

This work was partially supported by CAPES (Brazilian government
agency).


\end{document}